\documentclass[twocolumn,showpacs]{revtex4}
\usepackage{graphicx}
\usepackage{dcolumn}
\begin{document}

\title{Charge Ordering in the One-Dimensional Extended Hubbard
Model:\\ Implication to the TMTTF Family of Organic Conductors}

\author{Y. Shibata$^1$}
\author{S. Nishimoto$^1$}
\thanks{Present address: Many-Particle Theory Group, Department
of Physics, Philipps University Marburg, D-35032 Marburg, Germany.}
\author{Y. Ohta$^{1,2}$}
\affiliation{$^1$Graduate School of Science and Technology,
Chiba University, Chiba 263-8522, Japan}
\affiliation{$^2$Department of Physics, Chiba University,
Chiba 263-8522, Japan}
\date{4 June 2001}

\begin{abstract}
We study the charge ordering (CO) in the one-dimensional (1D)
extended Hubbard model at quarter filling where the nearest-neighbor
Coulomb repulsion and dimerization in the hopping parameters are
included.  Using the cluster mean-field approximation to take into
account the effect of quantum fluctuations, we determine the CO
phase boundary of the model in the parameter space at $T=0$ K.
We thus find that the dimerization suppresses the stability of
the CO phase strongly, and in consequence, the realistic parameter
values for quasi-1D organic materials such as (TMTTF)$_2$PF$_6$
are outside the region of CO.  We suggest that the long-range
Coulomb interaction between the chains should persist to stabilize
the CO phase.
\end{abstract}
\pacs{72.80.Le, 71.27.+a, 71.10.Fd, 71.30.+h,75.40.Mg}
\maketitle

\section{INTRODUCTION}

The charge-ordering (CO) phase transition has recently attracted
much attention in physics of strongly correlated electron
systems such as transition-metal oxides and organic charge-transfer
salts.  A possible mechanism of CO is the localization of electrons
on a lattice due to long-range Coulomb repulsions, which is a
lattice version of the Wigner crystallization.  The spin degrees
of freedom of the system of CO are yet active and can undergo an
additional phase transition by lowering temperature, to result in
a variety of magnetic ground states such as spin-Peierls (SP),
spin-density-wave (SDW), and antiferromagnetic (AF) states.
One of the simplest models that allow for such CO is the
one-dimensional (1D) Hubbard model with onsite ($U$) and
nearest-neighbor ($V$) Coulomb repulsions, where the charge density
wave (CDW) with two-fold periodicity, i.e., $4k_{\rm F}$-CDW with
$k_{\rm F}$ being the Fermi momentum, is realized when the band
filling is either 3/4 or 1/4 (quarter-filling).

Bechgaard salts (TMTSF)$_2$X and their sulfur analogs (TMTTF)$_2$X,
where X is PF$_6$, Br, ClO$_4$, etc., offer a series of materials
suitable for studying quasi-1D correlated electron systems at
quarter filling \cite{bourbonnais}.
It is known that the lattice dimerization of the systems causes
an alternation of hopping integrals and makes the system a Mott
insulator.  The metal-insulator transition of the series is then
controlled by the interchain hopping parameter $t_\perp$, i.e., by
the dimensional crossover from 1D to 2D
\cite{vescoli,suzumura,kishine}.

Recently, clear evidence of the CO phase transition in
(TMTTF)$_2$PF$_6$ has been given by the measurements of dielectric
response \cite{nad} and NMR spectroscopy \cite{chow}.  This
material has the smallest interchain coupling $t_\perp$ in
the series and is in the 1D confinement regime
\cite{bourbonnais,vescoli}.
The transition temperature of CO is reported to be
$T_{\rm CO}\simeq 100$ K, which is very high for small energy
scales of organic systems.  A new research area for studying
CO has thus been established \cite{chow}.

A number of theoretical studies have so far been made
on the 1D extended Hubbard model at quarter filling
\cite{milazotos,pencmila,sano,yoshioka,clay,nakamura,seo,ogata1,ogata2,tomio
},
which provide useful information to consider the physics of CO.
While some of them have not included the effect of lattice
dimerization \cite{milazotos,pencmila,sano,yoshioka,clay,nakamura},
recent mean-field calculations \cite{seo,ogata1,ogata2,tomio} have
taken into account the effect and explained how the $2k_{\rm F}$-
or $4k_{\rm F}$-CDW coexists with $2k_{\rm F}$-SDW in the ground
state of the systems.  We note however that the mean-field
analyses usually overestimate the stability of the ordered
phase.  Coupling to the lattice degrees of freedom has also
been argued to play an essential role \cite{mazumdar}.

Motivated by such development in the field, we study in this
paper the CO of the 1D extended Hubbard model with the
dimerization of hopping parameters, and consider its
implication to the observed CO in the TMTTF family of organic
conductors.
We use a new method of calculating the long-range order in
strongly-correlated electron models, i.e., an extension
of the so-called cluster mean-field approach \cite{bethe} to
correlated fermion systems.
Thus, the amplitude of CO, rather than its correlation
functions or exponents, can be calculated directly as in the
standard mean-field theory.  Moreover, critical interaction
strength of the quantum phase transition can be evaluated with
improved accuracy since the effect of quantum fluctuations is
automatically taken into account.  The CO phase boundary and
its dimerization dependence of the model is thus determined
in the parameter space $(U,V)$ at $T=0$ K.

We will thereby show that the critical interaction strength
$V_c$ is much larger than the value obtained in the mean-field
approximation, as expected, and also that the dimerization in
the hopping integrals suppresses the stability of the CO phase
strongly.
Then, it follows that the present model does not have the
CO ground state if we assume realistic values of the electronic
parameters for quasi-1D organic materials (TMTTF)$_2$PF$_6$
and (TMTSF)$_2$ClO$_4$.  This result suggests that the present
framework of the model under the influence of weak
three-dimensionality (3D) is not sufficient for describing the
CO observed in (TMTTF)$_2$PF$_6$.  Inclusion of any additional
degrees of freedom, in particular the 3D long-range Coulomb
interaction between the chains, may be essential for stabilizing
the CO phase.

\section{MODEL AND METHOD}

The extended Hubbard model is defined by the Hamiltonian
\begin{eqnarray}
H=&-&t_1\sum_{i=1}^{L/2}\sum_{\sigma}
(c_{2i-1,\sigma}^{\dagger}c_{2i,\sigma}+{\rm H.c})
\nonumber \\
&-&t_2\sum_{i=1}^{L/2-1}\sum_{\sigma}
(c_{2i,\sigma}^{\dagger}c_{2i+1,\sigma}+{\rm H.c.})
\nonumber \\
&+&U\sum_{i=1}^{L}n_{i,\uparrow}n_{i,\downarrow}
+V\sum_{i=1}^{L-1}n_{i}n_{i+1}
\end{eqnarray}
on the 1D lattice of size $L$ (even), where
$c_{i,\sigma}^\dagger$ ($c_{i,\sigma}$) is the electron
creation (annihilation) operator at site $i$ and spin $\sigma$
$(=\uparrow,\downarrow)$, and $n_i=n_{i,\uparrow}+n_{i,\downarrow}$
is the number operator.  We introduce a dimerization in the
hopping parameters, $t_1\ge t_2$.
$U$ and $V$ are the onsite and nearest-neighbor Coulomb
repulsions, respectively.
We restrict ourselves to the case of quarter-filling.
\begin{figure}
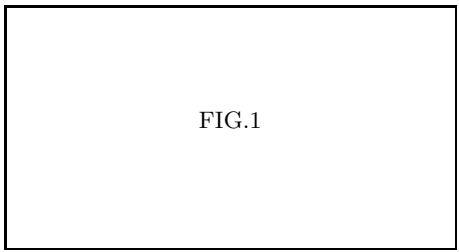

\begin{center}
\framebox[6cm]{\rule[-1.5cm]{0cm}{3cm}{FIG.1}}
\end{center}
\caption{Schematic representations of (a) the Hubbard chain
and (b) the chain with the mean-field bond.}
\label{fig:1}
\end{figure}

Our cluster mean-field method is the following:
We first introduce an exactly solved  finite-size system (or
a cluster, see Fig.~1) and then assume the presence of the
mean-field acting on the edges of the system.  The value of
the mean-field is then determined self-consistently so as
to minimize the total energy of the system.
In the present case, we prepare an $L$-site chain treated
exactly by a numerical method and assume that the mean-field
is applied on the edge sites $1$ and $L$ of the chain.
We then replace the mean-field by the `mean-field bond'
connecting between the sites $1$ and $L$; i.e., we assume
\begin{eqnarray}
n_1n_L\simeq n_1\langle n_L\rangle+\langle n_1\rangle n_L
-\langle n_1\rangle\langle n_L\rangle
\end{eqnarray}
for the bond, where $\langle n_1\rangle$ and $\langle n_N\rangle$
are the mean-fields at sites $i=1$ and $L$.
The Hamiltonian of the bond
\begin{eqnarray}
H_{1L}&=&-t_2\sum_\sigma(c_{1,\sigma}^\dagger c_{L,\sigma}
+{\rm H.c})
\nonumber \\
&+&V\big(n_1\langle n_L\rangle+\langle n_1\rangle n_L
-\langle n_1\rangle\langle n_L\rangle\big),
\end{eqnarray}
is then added to the Hamiltonian of the cluster Eq.~(1).
The total Hamiltonian is diagonalized numerically by the
Lanczos technique on small clusters, and the mean-fields are
evaluated as the expectation values of $n_1$ and $n_L$ for
the ground state.  The iterations are made to achieve
self-consistency in the values of the mean-fields
$\langle n_1\rangle$ and $\langle n_L\rangle$.
Converged solutions are obtained after $30-70$ iterations
depending on various conditions including the initial values
of $\langle n_1\rangle$ and $\langle n_L\rangle$.
We need to try a number of the initial values in order to
confirm that the iteration converges to the unique lowest-energy
solution.
In some cases, we find two (or possibly more) different
solutions, of which we choose the lowest-energy one; we should
note that the level crossing in the lowest-energy states
sometimes occurs in small-size systems (which is apparent,
e.g., in Fig.~3(a) given below).

\begin{figure}
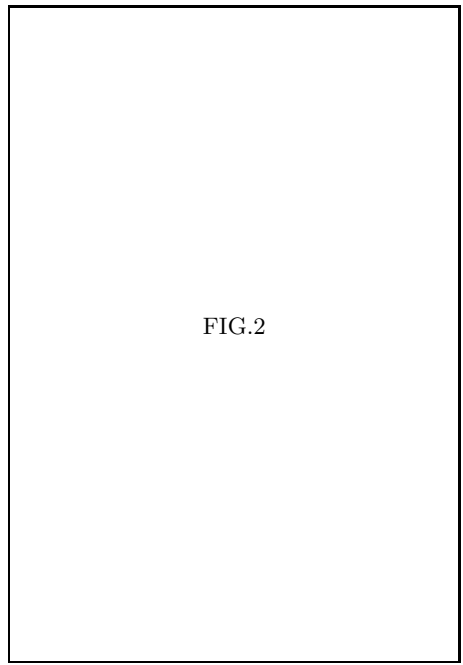

\begin{center}
\framebox[6cm]{\rule[-4.25cm]{0cm}{8.5cm}{FIG.2}}
\end{center}
\caption{Converged charge-density distributions for the models
without dimerization ($t_1$=$t_2$=$t$).  The cluster of length $L=8$
is used.  In (a), three typical solutions are shown: type I
(squares), type II (circles), and type III (triangles), which
are discussed in the main text.  In (b), oscillations of the
charge density near the critical point at $U/t=9$ are shown.}
\label{fig:2}
\end{figure}
In Fig.~2(a), we show the converged charge distributions for
three typical cases: the type I is for the paramagnetic
metallic state where a uniform charge distribution is seen,
the type II is for the CO state where the oscillation of
two-fold periodicity is clearly seen, and the type III is
for the parameter region corresponding to phase
separation \cite{sano,clay}.  In Fig.~2(b), we show how the
CO oscillation ceases near the critical point with decreasing
$V$; we find that the $4k_{\rm F}$ oscillation becomes
ill-defined between the interaction strength $V/t=2.67$ and
$2.68$, at which we decide the critical interaction strength
$V_c$ is located.  These are the cases without dimerization
but the situation is similar even when the dimerization
is introduced.  We should note that, in some cases, a
$2k_{\rm F}$ oscillation (corresponding to the effect of
Friedel impurity scattering) is superimposed on the $4k_{\rm F}$
CO oscillation, which is however rather small except in the
region of phase separation.

\section{RESULTS AND DISCUSSION}

\begin{figure}
\begin{center}
\framebox[6cm]{\rule[-6.25cm]{0cm}{12.5cm}{FIG.3}}
\end{center}
\caption{Dimerization $t_1/t_2$ dependence of the CO phase
boundary.  Results are shown for $t_1/t_2=1$ (squares),
$t_1/t_2=1.11$ (circles), and $t_1/t_2=1.43$ (triangles).
Clusters of the size (a) $L=8$, (b) $L=12$, and
(c) $L=16$ are used.  Upper-right side of the boundary is the
CO phase, and lower-left side of the boundary is either
paramagnetic metallic phase ($t_1=t_2$) or Mott insulating
phase ($t_1>t_2$).}
\label{fig:3}
\end{figure}
The obtained CO phase boundary of the model is shown in Fig.~3.
In case where there is no dimerization ($t_1=t_2=t$), we can
compare our results with those of the previous
studies \cite{sano,milazotos}, where the charge gap and
Luttinger-liquid parameter $K_\rho$ were evaluated in
small-size systems and the metal-insulator phase boundary
(which coincides with the CO phase boundary when $t_1=t_2$)
was determined.  We find that our results are in good agreement
with the previous results.  Exact solutions are available for
two limiting cases \cite{milazotos}:
$V_c/t=2$ at $U\rightarrow\infty$ and
$U_c/t=4$ at $V\rightarrow\infty$, which are also
consistent with our results.
We note that a rather inhomogeneous charge distribution
is obtained in the parameter region $V\gg U$ as seen
in Fig.~2(a); in this region, the model is reported to
have exotic phases where the superconducting pairing
fluctuations are dominant or the phase separation
occurs \cite{sano,clay}.

The cluster-size $L$ dependence of the critical interaction
strength $V_c$ for $t_1=t_2$ is shown in Fig.~4.
An oscillatory behavior corresponding to the so-called
shell effect for periodic boundary condition is found
between $L=8m$ (open shell) and $L=8m-4$ (closed shell)
where $m=1,2,\cdots$.  We find that the values of $V_c$ thus
calculated for $L=8$, 12, and 16 appear to be extrapolated
well to the value at $1/L\rightarrow 0$ which is estimated
from the density-matrix renormalization-group (DMRG) calculation
of the charge gap $\Delta_c$:  The results of $\Delta_c$ at
$1/L\rightarrow 0$ for several values of $V$ are fitted to
the exponential dependence \cite{tsuchiizu2}
$\Delta_c/t=\alpha\exp\lbrack-\beta t/(V-V_c)\rbrack$
(obtained from weak-coupling theory near the critical
point), from which we determine the value of $V_c$ (in
the accuracy of $\pm 0.05t$).
The details of our DMRG calculation is given in
Ref.\cite{nishimoto}
We expect from these results that the value of $V_c$
at $t_1>t_2$ also should not have too strong size-dependence
when $1/L\rightarrow 0$.
\begin{figure}
\begin{center}
\framebox[6cm]{\rule[-2.25cm]{0cm}{4.5cm}{FIG.4}}
\end{center}
\caption{Cluster-size $L$ dependence of the critical interaction
strength $V_c$.  Results are shown at $U/t=7$, $9$, and $12$
for the model without dimerization ($t_1=t_2=t$) .
The values of $V_c$ at $1/L\rightarrow 0$ determined from
the calculation of the charge gap by the density-matrix
renormalization group method are also shown for comparison.}
\label{fig:4}
\end{figure}

In case where the dimerization is present ($t_1>t_2$),
we first note that the charge gap opens \cite{nishimoto} except
for the noninteracting case ($U=0$); an infinitesimal $U$
value is enough to make the Umklapp process relevant in the
renormalization, leading to the opening of the charge gap.
When $U$ and $V$ becomes large, we find that the quantum
phase transition from this Mott insulating state to a CO
state occurs.  We can detect this transition by observing
the appearance of oscillations in the calculated charge
distribution.
Note that the mean-field bond is assumed to have the
hopping parameter $t_2$ which is smaller than $t_1$,
i.e., the bond is chosen so as to connect two `dimers'
(where the dimer is a pair of sites with the larger
hopping parameter).  This is because, in the strong
dimerization limit, only a single electron is present
in a dimer, and thus $V$ should not work in a dimer.

In Fig.~3, we find that the CO phase boundary exhibits
remarkably strong dimerization dependence; i.e., the
dimerization suppresses the stability of the CO phase.
General features are the following: (i) with increasing
dimerization, the boundary in the large $U$ region shifts
to the left, i.e., $V_c/t_2$ becomes large, and (ii) the
boundary in the large $V$ region shows a small upward
shift with increasing dimerization.
These features can be understood as follows:
We first note that, in the limit of strong dimerization
$t_1\gg t_2$, the critical $V$ value is determined as
$V_c=4t_1$ when $U\rightarrow\infty$.  This is because
in this limit the charge (which is either on the left
or on the right site of a dimer) may be expressed as the
pseudospin and the system becomes equivalent to the 1D
quantum Ising model \cite{sachdev}, the critical point of
which is determined by the competition between the quantum
fluctuation $t_1$ of an electron in a dimer and the
interaction $V$ between the two dimers leading to CO.
We therefore have $V_c/(4t_2)=t_1/t_2\rightarrow\infty$
when $t_1/t_2\rightarrow\infty$, which explains the behavior
(i).  When $V\rightarrow\infty$ and $U$ is finite, we find
from an estimation of kinetic energy of the lowest-energy
charge excitation that the critical $U$ value is given as
$U_c=2(t_1+t_2)$, irrespective of the strength of
dimerization.  Thus, we have $U_c/t_2=2(t_1/t_2+1)$,
which gives a contribution to the behavior (ii).  A clear
upward shift of the phase boundary is however not seen
in Fig.~3, which is partly because sufficiently large
values of $V$ and $t_1/t_2$ are not used and partly
because the finite-size effect becomes apparent here.

Besides the limiting cases discussed above, we can compare
our results with the results of the mean-field
calculation \cite{seo}.  Our results shown in Fig.~3 are for
two dimerization strengths, $t_2/t_1=0.9$ and $0.7$, which
correspond to the realistic values for (TMTSF)$_2$ClO$_4$
and (TMTTF)$_2$PF$_6$, respectively.  We then point out
that the critical interaction strength obtained here is
much larger than that of the mean-field calculation: e.g.,
we obtain $V_c/t_2\simeq 3$ at $U/t_2=5$ and $t_2/t_1=0.9$,
while the mean-field calculation \cite{seo} gives
$V_c/t_2\simeq 0.65$.  This is also the case without
dimerization $(t_1=t_2=t)$: e.g., we obtain $V_c/t=2.9$
at $U/t=5$, while the mean-field calculation \cite{seo}
gives $V_c/t=0.4$.  These discrepancies stem from the
strong quantum fluctuations of the present model.

Finally, let us consider experimental implication of our
results.  A recent careful estimation \cite{nishimoto} of
the values of the parameter $V$ for real materials has given
$V=0.21$ eV (or $V/t_2=0.8$) for (TMTSF)$_2$ClO$_4$ and
$V=0.18$ eV (or $V/t_2=2.0$) for (TMTTF)$_2$PF$_6$.  Also
the values $U/t_2=5.6$ for (TMTSF)$_2$ClO$_4$ and $U/t_2=10$
for (TMTTF)$_2$PF$_6$ have been reported \cite{nishimoto,mila}.
These values of the parameters in the $(U,V)$ plane
(see Fig.~3) are thus located in the Mott-insulating uniform
phase, far apart from the CO phase boundary.
We therefore conclude that the strength of the nearest-neighbor
Coulomb repulsion for these two materials is too small for
the corresponding 1D extended Hubbard model with dimerization
to be in the region of CO.  Since there is no CO at $T=0$ K,
it seems quite unlikely that the present model at $T>0$ can
have the CO even on the implicit assumption of the presence
of weak 3D coupling.
Then, to explain the CO observed in (TMTTF)$_2$PF$_6$, it
seems necessary to take into account any additional degrees
of freedom in our 1D model; a comparatively large value of
$T_{\rm CO}$ may suggest that there exists a rather strong
interchain coupling via the long-range Coulomb interaction,
which works to stabilize the CO phase.
We may say that, while the electron conduction of
(TMTTF)$_2$PF$_6$ at $T>T_{\rm CO}$ is of the 1D
nature \cite{bourbonnais,vescoli}, the observed CO is due
to the presence of the 2D--3D Coulombic coupling between the
chains.
Coupling to the anionic potential may also play an important
role in the stabilization of CO as has recently been suggested
in Ref.\cite{monceau}.
Whether these effects may be renormalized into the parameter
values of our 1D model is a highly nontrivial problem; the
issue is thus left for future work.

\section{CONCLUSION}

We have studied the CO in the 1D extended Hubbard
model at quarter filling by introducing the dimerization in the
hopping parameters as well as the nearest-neighbor Coulomb
repulsion.  We have used the cluster mean-field approximation
to take into account the effect of quantum fluctuations, and
have determined the CO phase boundary and its dimerization
dependence of the model in the $(U,V)$ plane at $T=0$ K.
We have thus found that the realistic parameter values for
(TMTTF)$_2$PF$_6$ and (TMTSF)$_2$ClO$_4$ are well outside the
region of CO.  We have argued that the present 1D model does
not provide a sufficient framework for describing the
CO observed in (TMTTF)$_2$PF$_6$.

\begin{acknowledgments}
We thank Prof. Y. Suzumura for tutorial lectures and
enlightening discussion.
This work was supported in part by Grants-in-Aid for
Scientific Research (Nos.~11640335 and 12046216) from the
Ministry of Education, Science, Sports, and Culture.
Computations were carried out at the computer centers of
the Institute for Molecular Science, Okazaki, and the
Institute for Solid State Physics, University of Tokyo.
\end{acknowledgments}

\end{document}